\documentclass[sigconf]{acmart}

\usepackage{pifont}
\usepackage{algorithmic}
\usepackage{algorithm}
\usepackage{multirow}
\usepackage{hyperref}

\AtBeginDocument{
  }

\copyrightyear{2025}
\acmYear{2025}
\setcopyright{cc}
\setcctype{by}
\acmConference[GLSVLSI '25]{Great Lakes Symposium on VLSI 2025}{June 30-July 2, 2025}{New Orleans, LA, USA}
\acmBooktitle{Great Lakes Symposium on VLSI 2025 (GLSVLSI '25), June 30-July 2, 2025, New Orleans, LA, USA}
\acmDOI{10.1145/3716368.3735145}
\acmISBN{979-8-4007-1496-2/2025/06}

\begin{document}

\title{Titanus: Enabling KV Cache Pruning and Quantization On-the-Fly for LLM Acceleration}

\author{Peilin Chen}
\email{peilin@virginia.edu}
\affiliation{
  \institution{University of Virginia}
  \city{Charlottesville}
  \state{VA}
  \country{USA}
}

\author{Xiaoxuan Yang}
\email{xiaoxuan@virginia.edu}
\affiliation{
  \institution{University of Virginia}
  \city{Charlottesville}
  \state{VA}
  \country{USA}
}

\begin{abstract}

Large language models~(LLMs) have gained great success in various domains. Existing systems cache Key and Value within the attention block to avoid redundant computations. However, the size of key-value cache~(KV cache) is unpredictable and can even be tens of times larger than the weights in the long context length scenario. In this work, we propose \textbf{Titanus}, a software-hardware co-design to efficiently compress the KV cache on-the-fly. We first propose the cascade pruning-quantization~(CPQ) method to reduce the KV cache movement. The hierarchical quantization extension strategy is introduced to tackle the non-independent per-channel quantization issue. To further reduce KV cache movement, we transfer only the non-zero KV cache between the accelerator and off-chip memory. Moreover, we customize a two-stage design space exploration framework for the CPQ method. A novel pipeline and parallelism dataflow is designed to reduce the first token generation time. Experiments show that \textbf{Titanus} achieves $159.9\times$~($49.6\times$) and $34.8\times$~($29.2\times$) energy efficiency~(throughput) compared to Nvidia A100 GPU and FlightLLM respectively. The code for \textbf{Titanus} is available at \href{https://github.com/peilin-chen/Titanus-for-LLM-acceleration}{this link}.

\end{abstract}



\maketitle

\section{Introduction}

Transformer-based large language models~(LLMs) have achieved state-of-the-art~(SOTA) results in a wide range of natural language processing tasks~\cite{yang2024harnessing,zhang2024systematic}. The attention mechanism, which combines the computation of query, key, and value, is an important building block for the LLMs. To improve the capabilities of LLMs, the model size has expanded from billions to trillions of parameters. Such large-scale LLMs consume significant amounts of energy to support their inference\cite{luo2024addition}. In order to reduce the parameter size, prior work~\cite{lin2024awq} quantizes LLM weights into low-bit. Moreover, prior works~\cite{xiao2023smoothquant, dettmers2022gpt3, shao2023omniquant, ashkboos2023towards} quantize both weights and activations to achieve significant improvements in latency and memory reduction. 

However, the key-value cache~(KV cache), which stores the attention keys and values to avoid redundant computations, is becoming the emerging memory bottleneck for LLM inference. For example, in the OPT-175B model~\cite{zhang2022opt}, with a batch size of 16, an input context length of 512, and an output context length of 1024, KV cache size can reach up to $1.2 TB$, which is $3.8\times$ the model's weights~\cite{sheng2023flexgen}. Most hardware platforms like GPUs and the SOTA transformer accelerator FlightLLM~\cite{zeng2024flightllm} store all weights and KV cache on high bandwidth memory~(HBM). Due to the autoregressive nature in the decode stage, these platforms repeatedly reload all the LLM weights and KV cache to generate the next token. A large and inevitable portion of energy is consumed in the data movement process. More importantly, the KV cache cannot be pre-quantized like weights because the key and value are dynamically generated during inference. Therefore, we are inspired to design a specialized accelerator that supports KV cache pruning and quantization on-the-fly to reduce both data movement and computation overhead.  

On the algorithm side, ThinK~\cite{xu2024think} focuses on key cache pruning and does not prune the most recent tokens and newly generated key cache. This approach cannot guarantee a high pruning ratio when the prompt is short and the output context length is large. KIVI~\cite{liu2024kivi} and KVQuant~\cite{hooper2024kvquant} quantize KV cache into 3-bit and even 2-bit with a slight accuracy drop, using group-wise per-channel quantization~(PCQ) for key cache and group-wise per-token quantization~(PTQ) for value cache. However, we observe that the computation and storage overhead of PTQ is significantly larger than PCQ due to the long input/output context length. Moreover, we find that KV cache across different layers, and even Key and Value within each layer, exhibit varying sensitivities to quantization bit-width. This motivates us to adopt a hardware-friendly and layer-sensitive mixed bit-width quantization for the KV cache, aiming to achieve significant compression with minimal hardware overhead.

On the hardware side, prior works $A^3$~\cite{ham20203}, ELSA~\cite{ham2021elsa}, DOTA~\cite{qu2022dota}, and FlightLLM~\cite{zeng2024flightllm} propose various techniques to address the quadratic computation and memory complexity challenge of self-attention mechanism, including approximate self-attention algorithm, low-rank linear transformation, and utilizing block-wise and N:M sparsity patterns. Previous hardware designs ignore the KV cache optimization and overlook the inefficiency of repeatedly transferring all weights and KV cache between the accelerator and memory during the decode stage. To tackle these challenges, we propose \textbf{Titanus}, which supports KV cache pruning and quantization on-the-fly and adopts chiplet-based computing-in-memory~(CIM)~\cite{yang2022research} design to eliminate static weights movement.

Specifically, the contributions of our work are as follows.
\begin{enumerate}
    \item We design the specialized hardware \textbf{Titanus} to support KV cache pruning and quantization on-the-fly to reduce the transferred KV cache size.
    \item We propose the cascade pruning-quantization compression method to reduce KV cache movement. Moreover, hierarchical quantization extension strategy is designed to address the non-independent per-channel quantization issue.
    \item To further reduce KV cache movement, we transfer only the non-zero KV cache between the accelerator and off-chip memory, omitting the zeros from pruning and quantization.
    \item We explore optimal pruning and quantization configurations for Key and Value in different LLM layers through a two-stage design space exploration~(DSE).
    \item We introduce the intra-core pipeline and inter-core parallelism dataflow to reduce the first token generation time.
\end{enumerate}
\vspace{-10pt}

\section{Preliminary and Motivation}

\subsection{Computing-in-Memory}

The computing-in-memory architecture has a significant energy efficiency advantage against the conventional platforms~(CPU, GPU, and FPGA)~\cite{song2017pipelayer,verma2019memory,yang2020retransformer}. Prior CIM designs~\cite{tu202228nm1, chih202116, tu202228nm2} have limited chip area and cannot store all weights of LLMs. The large-size data movement still exists in their CIM designs, which makes it challenging to achieve the best performance brought by CIM~\cite{chen2025optimizing}. 

In this work, to fully utilize the advantage of CIM architecture, we design the chiplet-based CIM accelerator to avoid repeatedly weight reloading by keeping all static weights on-chip. We do not store the dynamic weight KV cache on-chip due to the unpredictable size of the KV cache. To tackle the KV cache movement challenge, \textbf{Titanus} compresses the KV cache significantly on-the-fly to reduce dynamic weight movement while maintaining the accuracy of LLMs.

\subsection{Group-wise PCQ and PTQ} 

Prior algorithm works~\cite{liu2024kivi, hooper2024kvquant} quantize KV cache in both per-channel and per-token dimension using uniform quantization method~\cite{yang2022hero}. However, the overhead of PTQ is significantly larger than PCQ considering both the computation and storage overhead of quantization parameters~(including scale factor and zero point) due to the long input/output context length. For example, in OPT-125M model, with a total context length of 2048, the computation and storage overhead of PTQ can reach $32\times$ compared to PCQ. Moreover, to achieve lower-bit KV cache quantization, prior works~\cite{liu2024kivi, hooper2024kvquant} also adopt the group-wise PCQ~(GPCQ), i.e., group elements along the channel dimension and quantize them together. Compared to PCQ, GPCQ generates a large number of quantization parameters because it needs to store a new set of scale factor and zero point for each group. Based on our experimental results, the transferred KV cache and quantization parameters of PCQ are even smaller than those of GPCQ, while PCQ has higher accuracy than GPCQ. Therefore, the PCQ is more hardware-friendly than PTQ and GPCQ.
\vspace{-16pt}

\subsection{Non-independent Per-channel Quantization}
\label{section_2.3}

Non-independent per-channel quantization~(NiPCQ) issue refers to the need to re-quantize all channels whenever a newly generated token arrives. The NiPCQ issue will lead to much additional computation overhead because the prior tokens are quantized repeatedly. A straightforward approach adopted by prior algorithm work~\cite{liu2024kivi} is to group the KV cache based on the group size and quantize them separately. However, this method has two drawbacks. One is the data movement and storage overhead of GPCQ, as discussed in the previous subsection. The other is the large on-chip buffer requirement to keep the group KV cache in full precision. In our work, we propose the Hierarchical Quantization Extension strategy to address this challenge efficiently.
\vspace{-5pt}

\section{Algorithm Design}

\subsection{KV Cache Cascade Compression}

\label{section_3.1}

\begin{figure}[tb]
    \centering
    \includegraphics[width=\linewidth]{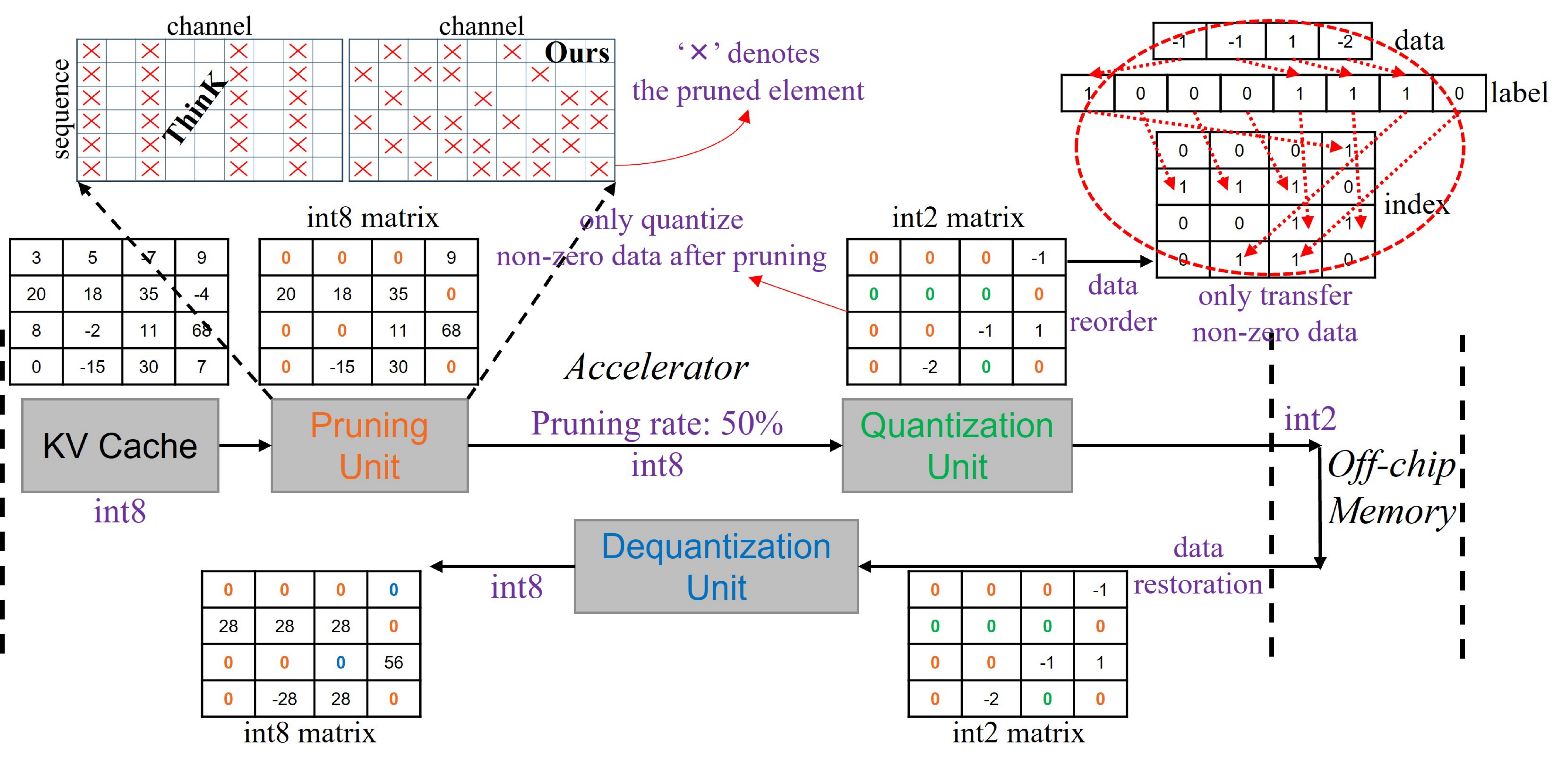}
    \vspace{-10pt}
    \caption{Overview of cascade pruning-quantization method.} 
    \Description{fig1}
    \label{fig1}
    \vspace{-10pt}
\end{figure}

We leverage the benefits of pruning and quantization and combine the compression effects, and thus the pruning is set as the first step. Following that, the (de)quantization process will not impact the pruned KV cache elements. Therefore, our cascade pruning-quantization~(CPQ) is effective in reducing computation overhead.

For the pruning part, prior algorithm work ThinK~\cite{xu2024think} prunes the KV cache at the channel granularity and overlooks the most recent tokens and newly generated KV cache. With their method, it is difficult to achieve a high pruning ratio while maintaining the LLM accuracy since pruned channels may have a negative impact on the specific token. Instead, we find it necessary to adopt a finer granularity strategy, which means that we only prune insignificant elements within the KV cache. An element is considered unimportant if its absolute value is below the given pruning threshold. The differences between ThinK and our method are shown in the upper-left side of Figure~\ref{fig1}. Moreover, we prune not only the prefill stage KV cache but also the dynamically generated KV cache. Note that, even though our pruning method has an unstructured pattern, our hardware design can handle this type of sparsity efficiently, as explained in Section~\ref{section_4.2}.

Figure~\ref{fig1} depicts the overall CPQ compression method. The int8 KV cache first goes through the pruning unit~(PU). PU generates the pruned KV cache and binary index that marks the position of non-zero data. The quantization unit~(QU) only quantizes the non-zero data in the pruned KV cache and generates the binary label to mark the position of the non-zero data after quantization. There are more zero elements after quantization~(green elements shown in the int2 matrix). Then, we reorder the quantized KV cache and only transfer the non-zero elements and auxiliary data~(index and label). The dequantization unit will use the auxiliary data to reconstruct the quantized KV cache. Therefore, we can significantly reduce the KV cache movement. Moreover, our experiments demonstrate that there are more zero elements generated from the dequantization step. We leverage the zero data from pruning and dequantization together to reduce computation overhead. Note that we quantize the int8 KV cache into lower-bit here since prior work~\cite{xiao2023smoothquant} has verified that 8-bit precision for weight and activation~(W8A8) can achieve almost no accuracy loss and \textbf{Titanus} mainly supports integer computation.
\vspace{-5pt} 

\begin{figure}[tb]
    \centering
    \includegraphics[width=\linewidth]{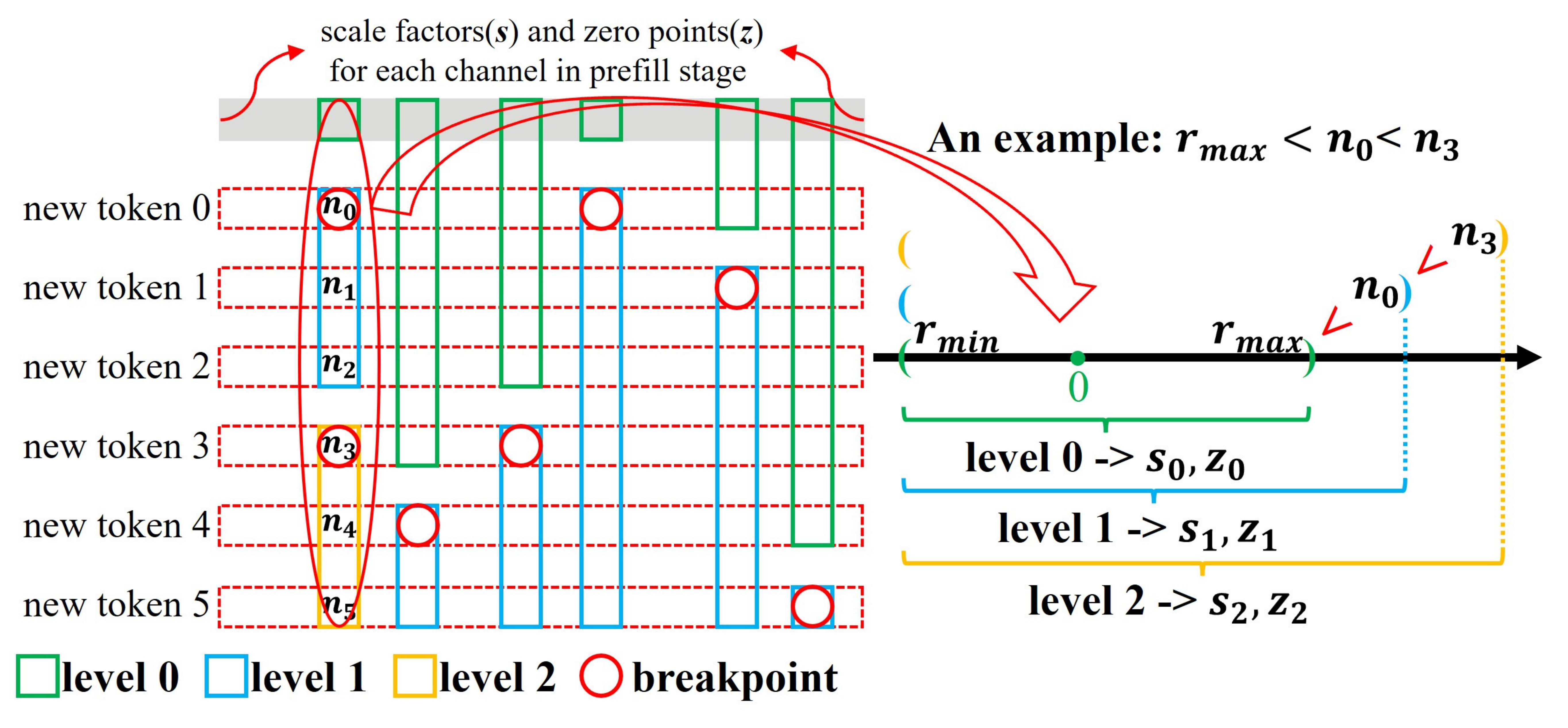}
    \vspace{-10pt}
    \caption{Hierarchical quantization extension strategy.}
    \Description{fig2}
    \label{fig2}
    \vspace{-10pt}
\end{figure}

\subsection{Hierarchical Quantization Extension (HQE)}
\label{section_3.2}

In order to tackle the NiPCQ challenge~(in Section~\ref{section_2.3}) and avoid the large overhead, we propose the hierarchical quantization extension strategy. HQE divides the channel into multiple levels and extends the tolerance range~(TR) for the high levels, which guarantees that each token is quantized only once with little storage overhead. In Figure~\ref{fig2}, $r_{max}$ and $r_{min}$ denote the maximum and minimum data~(or TR) within the prefill stage. We first obtain the scale factors and zero points for each channel in the prefill stage. Then, we design a monitoring mechanism and calculate the TR using the parameters~($r_{max}$, $r_{min}$, and $s_0$) obtained in the prefill stage. The newly generated token during the decode stage will be checked whether the data is within the TR. If the new token is out of range, we generate a higher level for future tokens in the decode stage and extend the TR simultaneously. Otherwise, the new token belongs to the prior level. As shown in Figure~\ref{fig3}, the breakpoint $n_{0}$ generates the new level whereas $n_{1}$ and $n_{2}$ maintain the same level as $n_{0}$. When a new level is generated, we save the scale factor and zero point and extend the TR. The HQE method reuses the scale factor and zero point effectively, thus resulting in negligible storage overhead for quantization parameters.

\vspace{-6pt}

\subsection{Two-Stage Design Space Exploration}

\begin{figure}[tb]
    \centering
    \includegraphics[width=\linewidth]{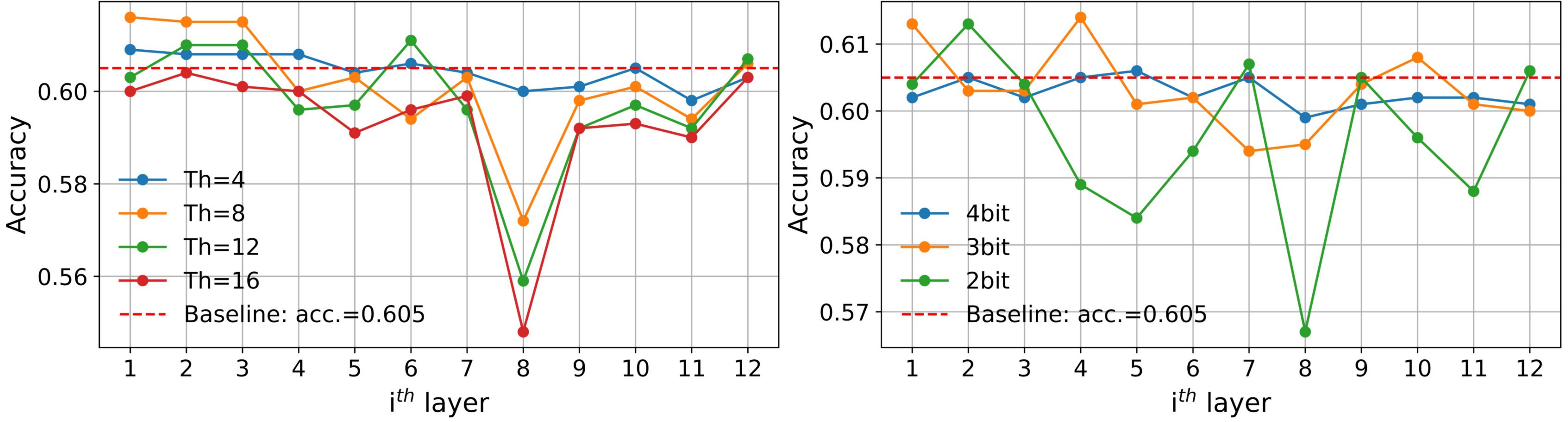}
    \vspace{-10pt}
    \caption{Sensitivity of KV cache across different layers to pruning threshold~(left) and quantization bit-width~(right).}
    \Description{fig3}
    \label{fig3}
    \vspace{-10pt}
\end{figure}

To explore the optimal KV cache pruning and quantization configurations, we need to understand the sensitivity of KV cache across different layers. Our experiment shows that KV cache across different layers exhibits varying sensitivities to the pruning threshold~(Th) and quantization bit-width. Moreover, Key and Value also exhibit different sensitivities. Figure~\ref{fig3} presents the KV cache sensitivity result of OPT-125M model on the LAMBADA dataset~\cite{paperno2016lambada}. We observe that layer 8 is the most sensitive for pruning and uniform 2-bit quantization may not work due to the limited data range.
We customize a two-stage DSE framework for our CPQ compression method. We first leverage the multi-objective optimization algorithm NSGA-II~\cite{deb2002fast} to find the Pareto optimal KV cache pruning configuration. The objectives here are accuracy and average pruning ratio. However, NSGA-II is not ideal for quantization configuration search since quantizing the KV cache can be $20\times$ more time-consuming than pruning. To address this issue, we design the second DSE stage. Firstly, we quantize the KV cache by progressively reducing the quantization bit-width~(from 8-bit to 2-bit). Secondly, we visualize the KV cache distribution and check whether the Key or Value is more sensitive to quantization. Thirdly, we increase the quantization bit-width for all layers of the more sensitive part. Finally, we further adjust the quantization bit-width for less sensitive part in certain layers based on the extrema distribution. We can leverage the above workflow to shrink the search space and rapidly determine the Pareto-optimal quantization bit-width for the KV cache across different layers.
\vspace{-6pt}

\section{Hardware Architecture}
\label{section4}

\begin{figure*}[tb]
    \centering
    \includegraphics[width=0.87\linewidth]{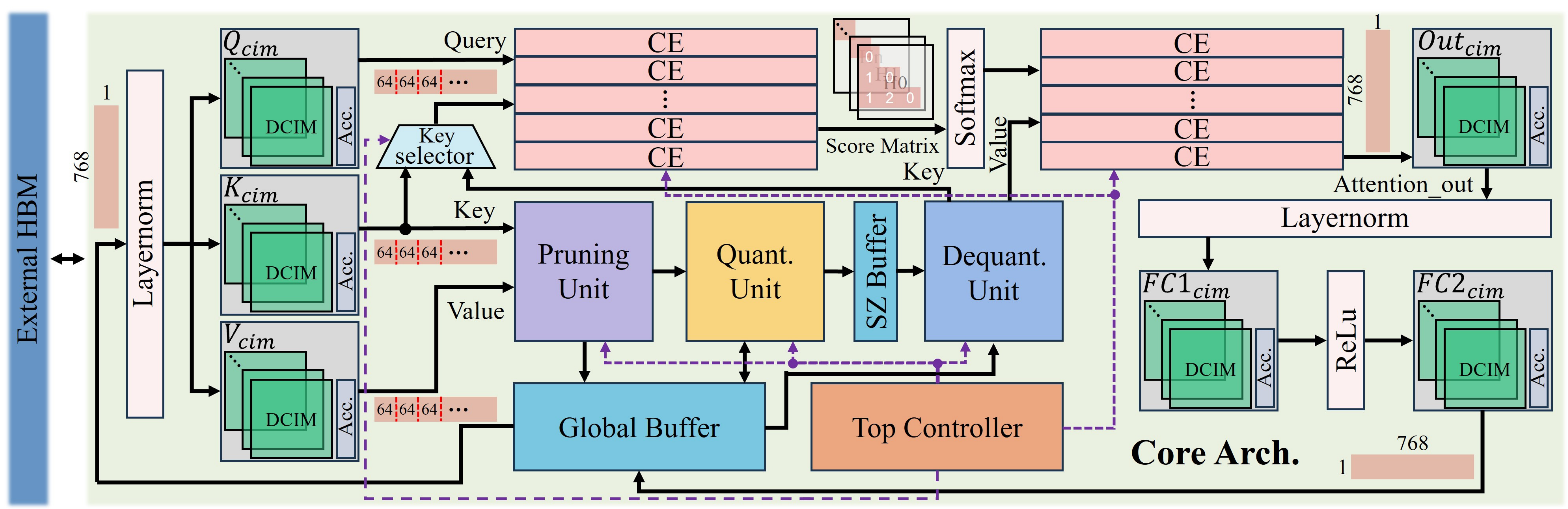}
    \caption{Titanus core-level overall architecture. CE and SZ denote the computing engine and scale-zero buffer, respectively.}
    \Description{fig4}
    \label{fig4}
    \vspace{-10pt}
\end{figure*}

\subsection{Overview}

Figure~\ref{fig4} depicts the \textbf{Titanus} core-level overall architecture, which is designed to support one OPT decoder layer. To leverage weight stationary dataflow, we utilize a multi-core approach for the LLM inference. We use digital CIM~(DCIM) macros~\cite{chih202116} and accumulators to construct the large CIM block. $Q_{cim}$, $K_{cim}$, $V_{cim}$, $Out_{cim}$, $FC1_{cim}$, and $FC2_{cim}$ blocks are designed to store six trainable matrices within the decoder layer. Importantly, we fuse multi-head Query, Key, and Value matrices into three separate large matrices and store them in the $Q_{cim}$, $K_{cim}$, and $V_{cim}$ block, respectively. We also design a novel computing engine~(CE) to exploit zeros from pruning and dequantization to reduce computation overhead. The CE number is consistent with the number of self-attention heads, which can process all heads in parallel. Pruning unit~(PU), quantization unit~(QU), and dequantization unit~(DQU) are designed to support the proposed CPQ KV cache compression method. Top controller guarantees that \textbf{Titanus} starts to calculate the score matrix after $Q_{cim}$, $K_{cim}$, and $V_{cim}$ blocks complete one token computation in the prefill stage. Thus, we first compute the diagonal elements in the score matrix. The scale-zero~(SZ) buffer is designed to store all the quantization parameters.
\vspace{-5pt}

\subsection{Computing Engine}
\label{section_4.2}

\begin{figure}[tb]
    \centering
    \includegraphics[width=\linewidth]{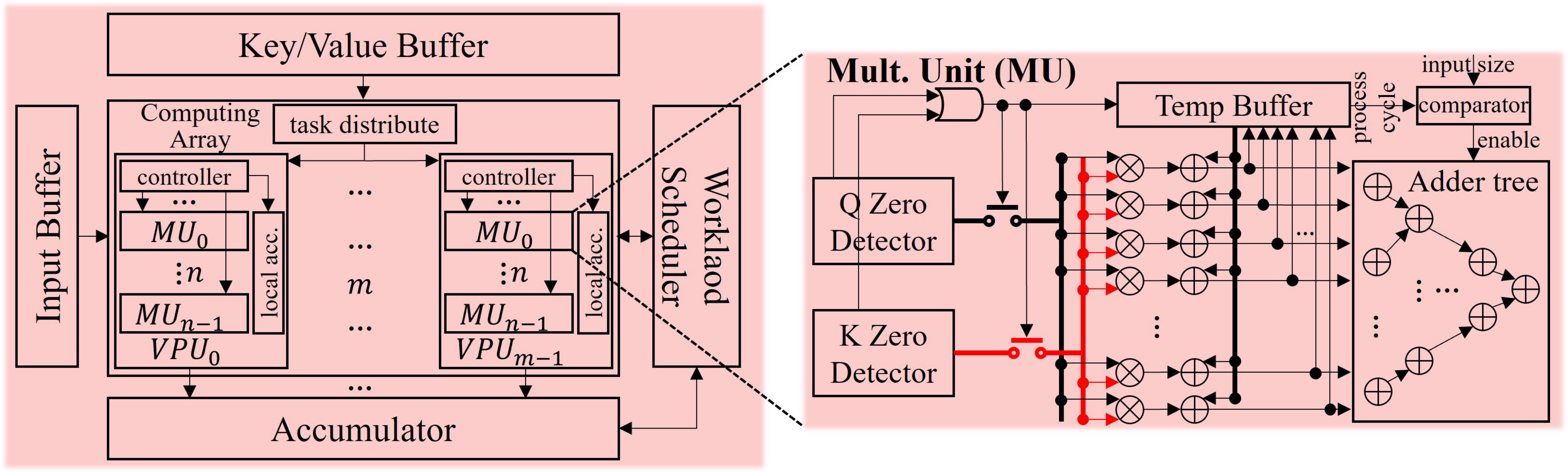}
    \vspace{-10pt}
    \caption{Computing engine design for dot-product attention with zero detection functionality.}
    \Description{fig5}
    \label{fig5}
    \vspace{-10pt}
\end{figure}

We design a novel CE that utilizes the irregular KV cache sparsity pattern to save energy. CE uses zero detector to detect and skip redundant computations and dynamically activates the required units based on the input vector size. As shown in Figure~\ref{fig5}, the CE module consists of computing array~(CA), workload scheduler~(WS), accumulator, and buffer. The CA activates all vector processing units~(VPUs) and evenly distributes tasks to the VPUs. The controller within each VPU can dynamically determine how many multiplication units~(MUs) to enable. Subsequently, the local accumulator aggregates the results from the enabled MUs. The MU can skip zero multiplications using zero detector, and thus reducing energy consumption. The comparator inside the MU controls the enable signal for the adder tree module based on the processed data size. The WS is designed to support three computation cases: (1) \textit{input vector size} $<$ \textit{CE max capability}, (2) \textit{input vector size} $=$ \textit{x}$\times$\textit{(CE max capability)}, and (3) \textit{input vector size} $=$ \textit{(x+y)}$\times$\textit{(CE max capability)}, where \textit{x} represents positive integer and $0<y<1$.
\vspace{-6pt}

\subsection{Pruning Unit}

\begin{figure}[tb]
    \centering
    \includegraphics[width=0.6\linewidth]{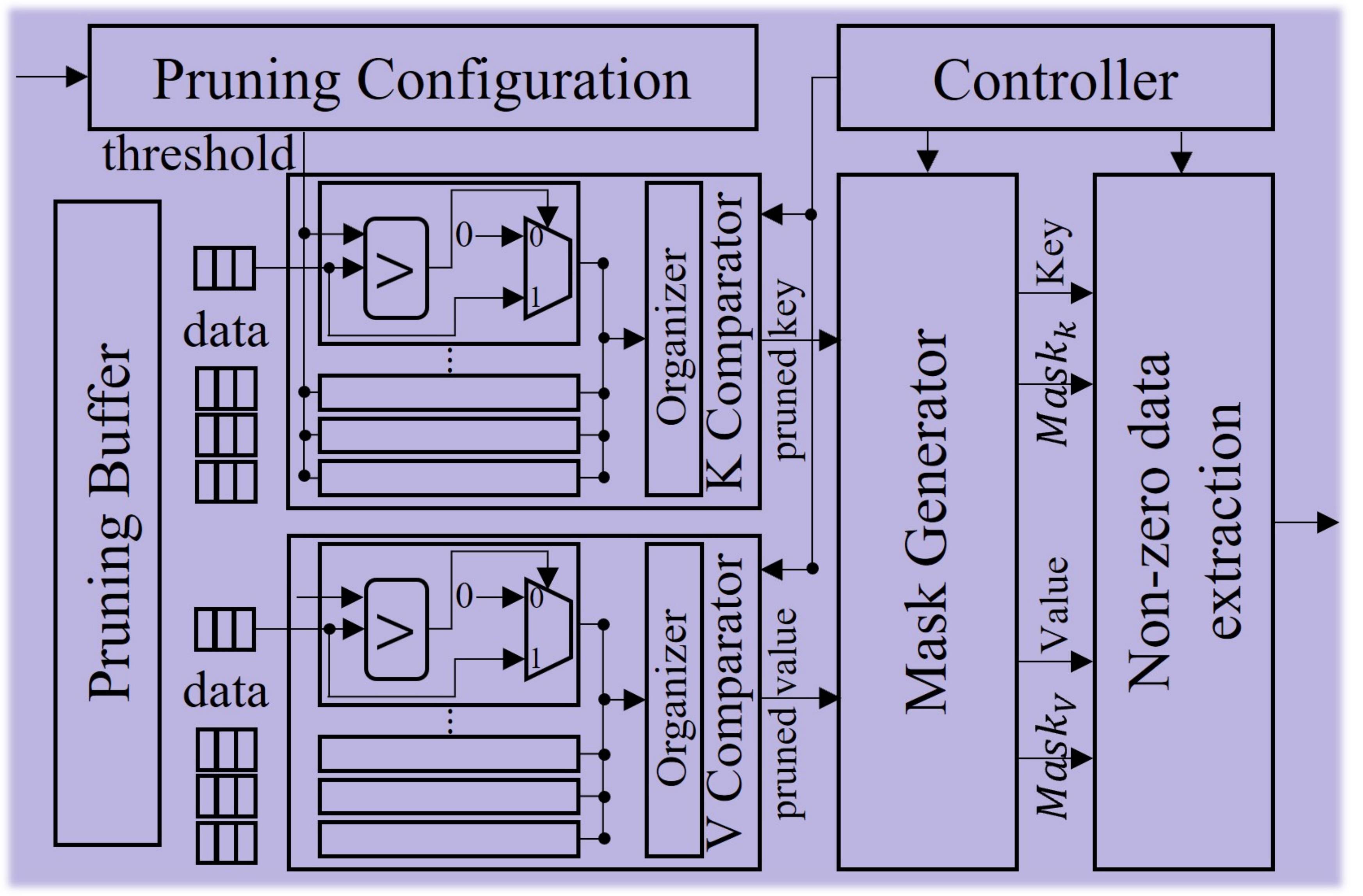}
    \vspace{-10pt}
    \caption{On-the-fly pruning unit design.}
    \Description{fig6}
    \label{fig6}
    \vspace{-10pt}
\end{figure}

To support the proposed pruning method in Section~\ref{section_3.1}, we design a specialized PU module that prunes the KV cache on-the-fly and only transfers the non-zero KV cache in the subsequent dataflow. In Figure~\ref{fig6}, the pruning configuration contains a list of thresholds for LLM's Key and Value across all layers, such as [$Th_{k0}$, $Th_{k1}$, $Th_{k2}$, ..., $Th_{v0}$, $Th_{v1}$, $Th_{v2}$, ...]. There are two comparators inside the PU module. Thus, Key and Value can be processed in parallel. The K comparator first fetches threshold data from the pruning configuration and then prunes the Key cache in parallel using multiple units. The controller dynamically determines the number of units within the K comparator to enable based on the input data size. Similarly, the V comparator adopts the same design. The mask generator produces the binary index that marks the position of non-zero data based on the pruned Key and Value. The non-zero data extraction module ensures that only the non-zero KV cache is transferred to the subsequent dataflow after pruning.
\vspace{-6pt}

\subsection{Quantization and Dequantization Unit}

\begin{figure}[tb]
    \centering
    \includegraphics[width=0.85\linewidth]{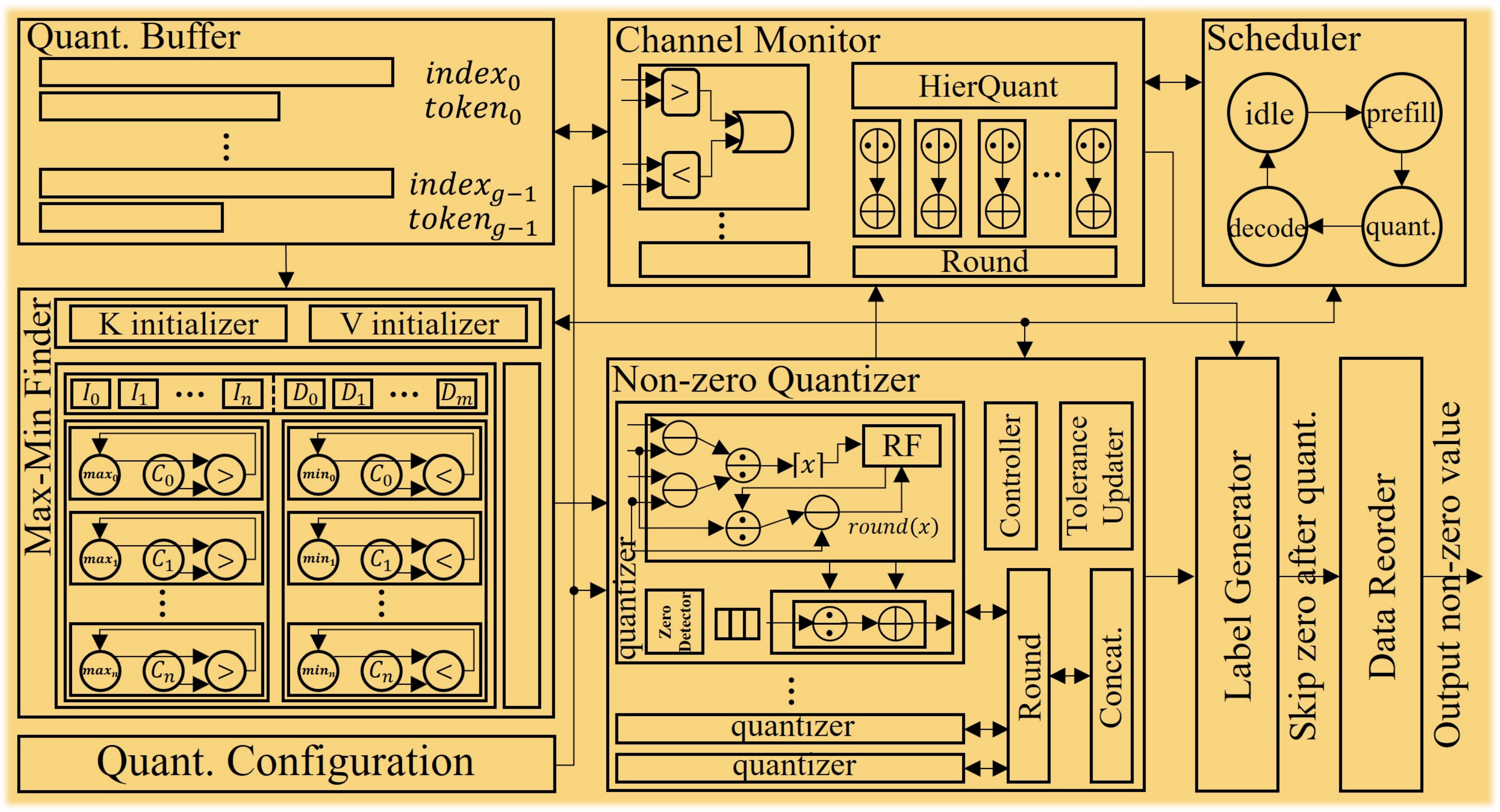}
    \vspace{-10pt}
    \caption{Quantization unit enabling the HQE method.}
    \Description{fig7}
    \label{fig7}
    \vspace{-10pt}
\end{figure}

We design a QU to support the proposed HQE method in Section~\ref{section_3.2}. The QU is more complex than CE and PU modules due to more building blocks and signal connections, as shown in Figure~\ref{fig7}. Quantization~(quant.) buffer stores the binary index and pruned non-zero KV cache from the PU. Quantization configuration stores the quantization bit-width for Key and Value across all layers. The Max-Min Finder~(MMF) computes the max and the min values for all channels of KV cache simultaneously in the prefill stage. K/V initializer in MMF sets max and min values to $-128$ and $127$. $I$, $D$, and $C$ denote the index of one token, the non-zero token data, and the current data. $n$ and $m$ represent the length of the input index and token~($n\geq m$). MMF updates the max and min values only when the index $I$ is 1~(non-zero data). The non-zero quantizer~(NZQ) is used in the prefill stage and quantizes Key and Value simultaneously. The quantizer within the NZQ calculates the scale factor and zero point for each channel and only quantizes the non-zero KV cache. The tolerance updater then computes the tolerance range for all channels based on the results from all quantizers. The channel monitor~(CM) checks whether the newly generated token is within the tolerance range and quantizes the KV cache in the decode stage. The HierQuant module inside CM can update the scale factor and zero point for higher levels locally. The scheduler uses finite state machine~(four states here) to organize the workflow of QU. The prefill, quant., and decode states control the MMF, NZQ, and CM modules, respectively. Finally, the QU only outputs the non-zero quantized KV cache and auxiliary data~(corresponding to Figure~\ref{fig1}).

DQU follows a design strategy similar to QU. There are three crucial points of the DQU. Firstly, the DQU utilizes the index and label to reconstruct the compressed KV cache. Secondly, the DQU does not dequantize the zero elements obtained from the pruning step~(i.e., $index==0$). Thirdly, the DQU does not use the subtractor and multiplier to perform dequantization if quantized data is either zero~(i.e., $label==0$) or equal to the \textit{zero point}.
\vspace{-6pt}

\subsection{Dataflow Design}

\begin{figure}[tb]
    \centering
    \includegraphics[width=\linewidth]{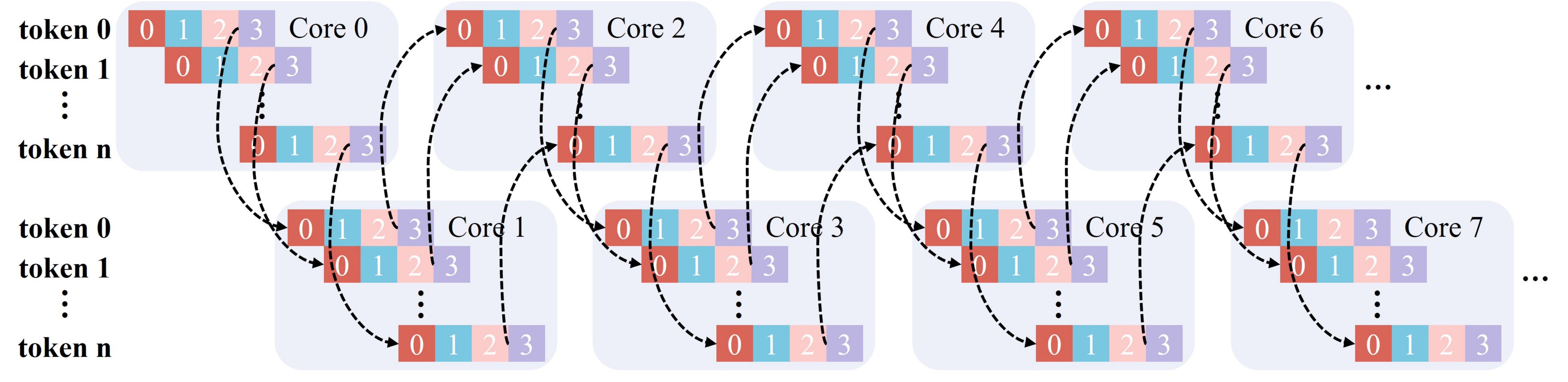}
    \vspace{-10pt}
    \caption{Intra-core pipeline and inter-core parallelism.}
    \Description{fig8}
    \label{fig8}
    \vspace{-10pt}
\end{figure}

For long input context length, the first token generation during the prefill stage is slow~\cite{fu2024lazyllm, horton2024kv}. To reduce the time-to-first-token~(TTFT), we propose the intra-core pipeline~(ICPI) and inter-core parallelism~(ICPA) dataflow design. As shown in Figure~\ref{fig8}, we divide the computation flow within the core-level into multiple stages. All prefill tokens are processed in a pipeline manner inside each core. Moreover, the $i^{th}$ core passes its result to the ${(i+1)}^{th}$ core after completing the computation of each token, thereby avoiding waiting for all tokens to be fully processed. Due to the data dependencies, the decode stage cannot be parallelized. Thus, the ICPI and ICPA only focus on reducing the TTFT. 

\section{Experimental Results and Analysis}

\subsection{Experimental Setup}

To provide a comprehensive evaluation of the proposed algorithm design, we choose five models of different sizes from the OPT family~\cite{zhang2022opt}, including OPT-125M, OPT-1.3B, OPT-2.7B, OPT-6.7B, and OPT-13B. The evaluation is performed on two zero-shot datasets: LAMBADA~\cite{paperno2016lambada} and TruthfulQA~\cite{lin2021truthfulqa}. We utilize the SmoothQuant~\cite{xiao2023smoothquant} to complete the W8A8 LLM quantization. All the above experiments are conducted on the NVIDIA A40~(48GB) and A100~(80GB) GPUs. 

We design the \textbf{Titanus} core-level architecture at register-transfer level~(RTL) using Verilog. We synthesize the \textbf{Titanus} using Synopsys Design Compiler~(DC) under a 14nm FinFET technology library~\cite{tuli2023acceltran} to estimate the area and power of the core modules. The area and power of the digital CIM macro are obtained from \cite{chih202116}. The area, power, and bandwidth of on-chip SRAM buffers~(\textit{Global} and \textit{SZ}) are estimated through CACTI~\cite{muralimanohar2009cacti}. We utilize NVMain~\cite{poremba2015nvmain} to simulate the off-chip HBM. According to the DC synthesis results, the critical path latency is less than 5ns. Therefore, we assume that \textbf{Titanus} operates at a frequency of 200MHz. Moreover, we build a cycle-accurate simulator to estimate the overall performance of \textbf{Titanus} based on AccelTran~\cite{tuli2023acceltran}. The code for \textbf{Titanus} is available at \url{https://github.com/peilin-chen/Titanus-for-LLM-acceleration}.

\begin{table}[tb]
\centering
\caption{ROUGE-1 score of baseline and HQE for OPT models.}
\label{table1}
\begin{tabular}{lccccc}
\toprule
\textbf{Model} & \textbf{125M} & \textbf{1.3B} & \textbf{2.7B} & \textbf{6.7B} & \textbf{13B} \\
\midrule
\multirow{1}{*}{Original Model} & 0.3983 & 0.4303 & 0.4311 & 0.4486 & 0.4442 \\
\midrule
\multirow{1}{*}{HQE} & 0.3989 & 0.4276 & 0.4289 & 0.4483 & 0.4392 \\
\bottomrule
\end{tabular}
\vspace{-7pt}
\end{table}

\begin{figure}[tb]
    \centering
    \includegraphics[width=\linewidth]{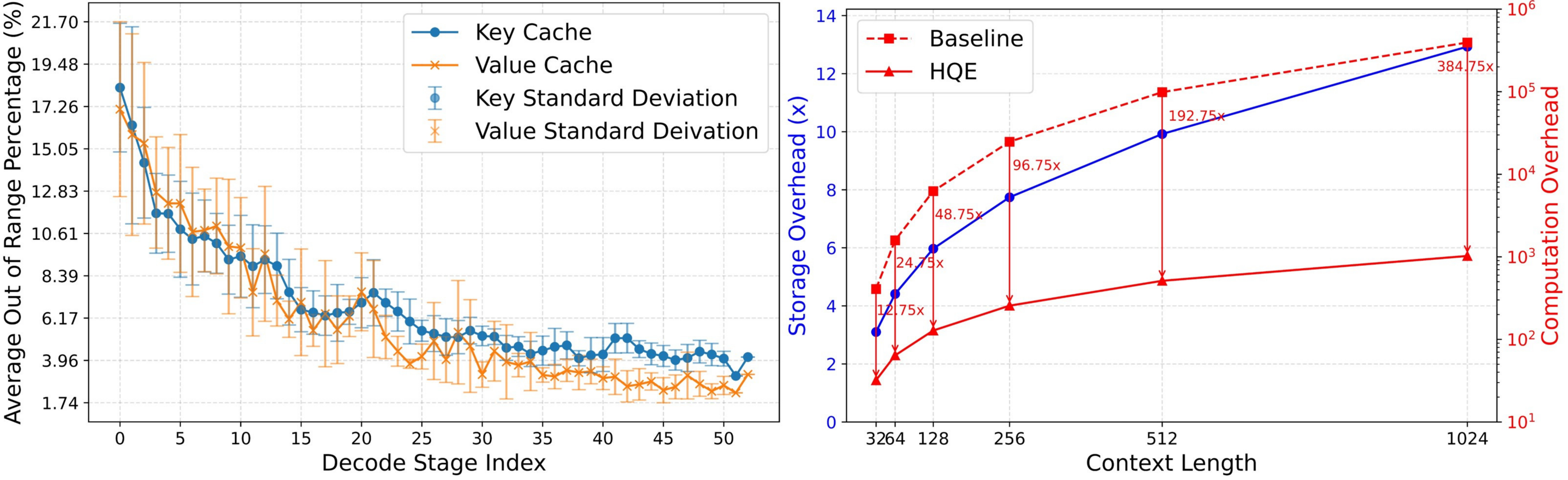}
    \vspace{-10pt}
    \caption{Left: Out of range percentage for KV cache across DS. Right: The storage overhead and computation reduction brought by HQE.}
    \Description{fig9}
    \label{fig9}
    \vspace{-6pt}
\end{figure}

\subsection{Algorithm Performance}

\textbf{Effectiveness of HQE Strategy.} We evaluate the HQE strategy using five OPT models under the generation task of TruthfulQA. In this experiment, we quantize 32-bit floating point~(fp32) KV cache into 3-bit~(int3). As shown in Table~\ref{table1}, the ROUGE-1 score~\cite{lin2004rouge} of HQE shows negligible degradation compared to the original model. When evaluating OPT-125M, HQE even achieves a higher {ROUGE-1} score than the original model. This method ensures that each token is quantized only once, avoiding repeatedly re-quantization of prior tokens when a newly generated token arrives. As shown in the right side of Figure~\ref{fig9}, HQE reduces quantization overhead by $384.75\times$ when the context length is 1024. Moreover, we present the out-of-range percentage for the OPT-125M KV cache using our HQE strategy in Figure~\ref{fig9}, where the maximum context length is 64. The out-of-range percentage gradually converges to 0 as the decode stage index increases. Thus, HQE introduces negligible storage overhead for quantization parameters due to the limited hierarchies in each channel. For example, the storage overhead of HQE is less than $13\times$ when the context length is 1024, which means that HQE only generates less than 13 levels per channel on average. 

\begin{figure}[tb]
    \centering
    \includegraphics[width=\linewidth]{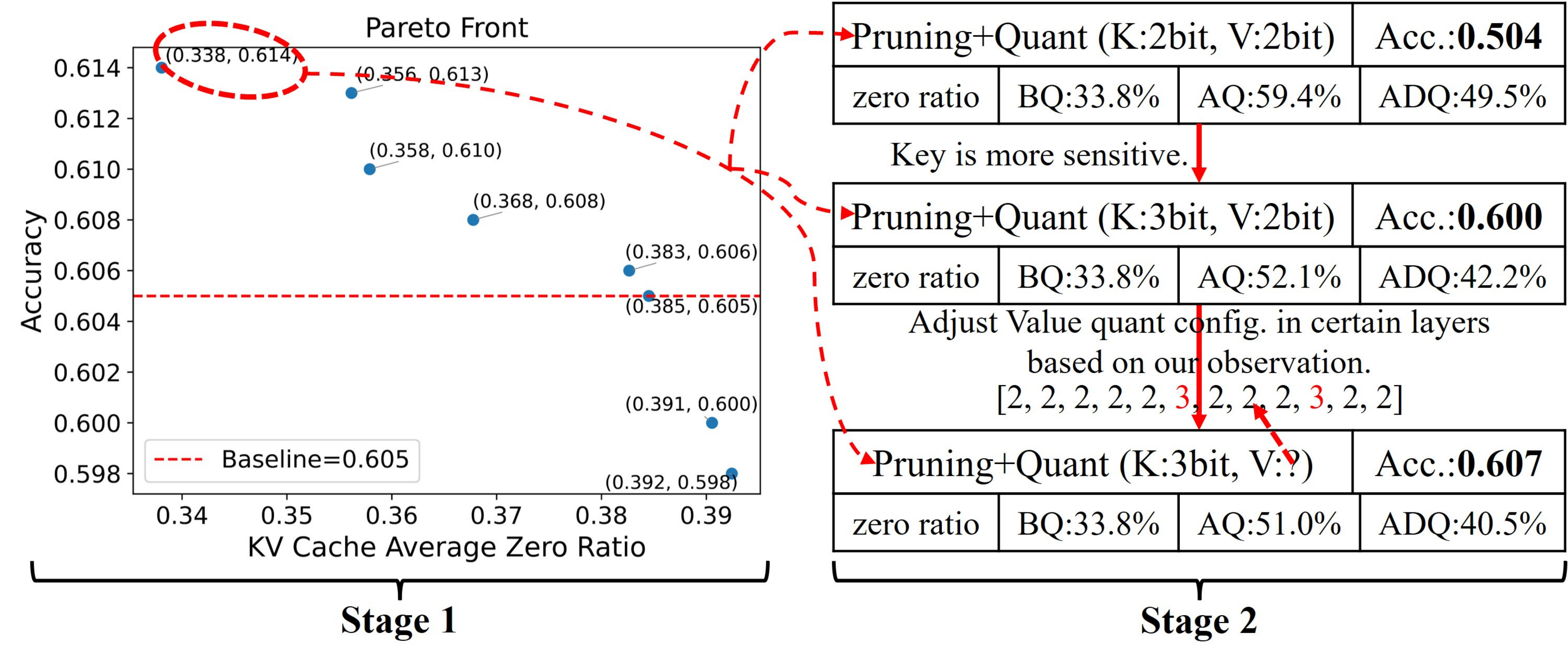}
    \vspace{-10pt}
    \caption{Two-stage DSE for the CPQ compression method. BQ: before quant. AQ: after quant. ADQ: after dequant.}
    \Description{fig10}
    \label{fig10}
    \vspace{-10pt}
\end{figure}

\textbf{Two-stage DSE for Key and Value.} We show the process to explore the optimal KV cache pruning and quantization configurations for OPT-125M in Figure~\ref{fig10}. The dataset is LAMBADA in this experiment. We first utilize NSGA-II to find the Pareto optimal pruning configuration. The search space is up to $24^5$, given 12 layers in OPT-125M, two variables~(Key and Value) per layer, and five possible options for each variable. We determine the potential options through experiments that analyze the accuracy loss under varying thresholds. Then, we select the Pareto solution with the highest accuracy to perform 2-bit-key and 2-bit-value quantization. We observe that Key is more sensitive to quantization. We increase the bit-width of Key and perform 3-bit-key and 2-bit-value quantization. To further restore the accuracy, we adjust the quantization bit-width from 2-bit to 3-bit for the Value cache in some layers where it is difficult to quantize. As a result, we can reduce the KV cache movement significantly while maintaining the LLM accuracy.   

\begin{figure}[tb]
    \centering
    \includegraphics[width=\linewidth]{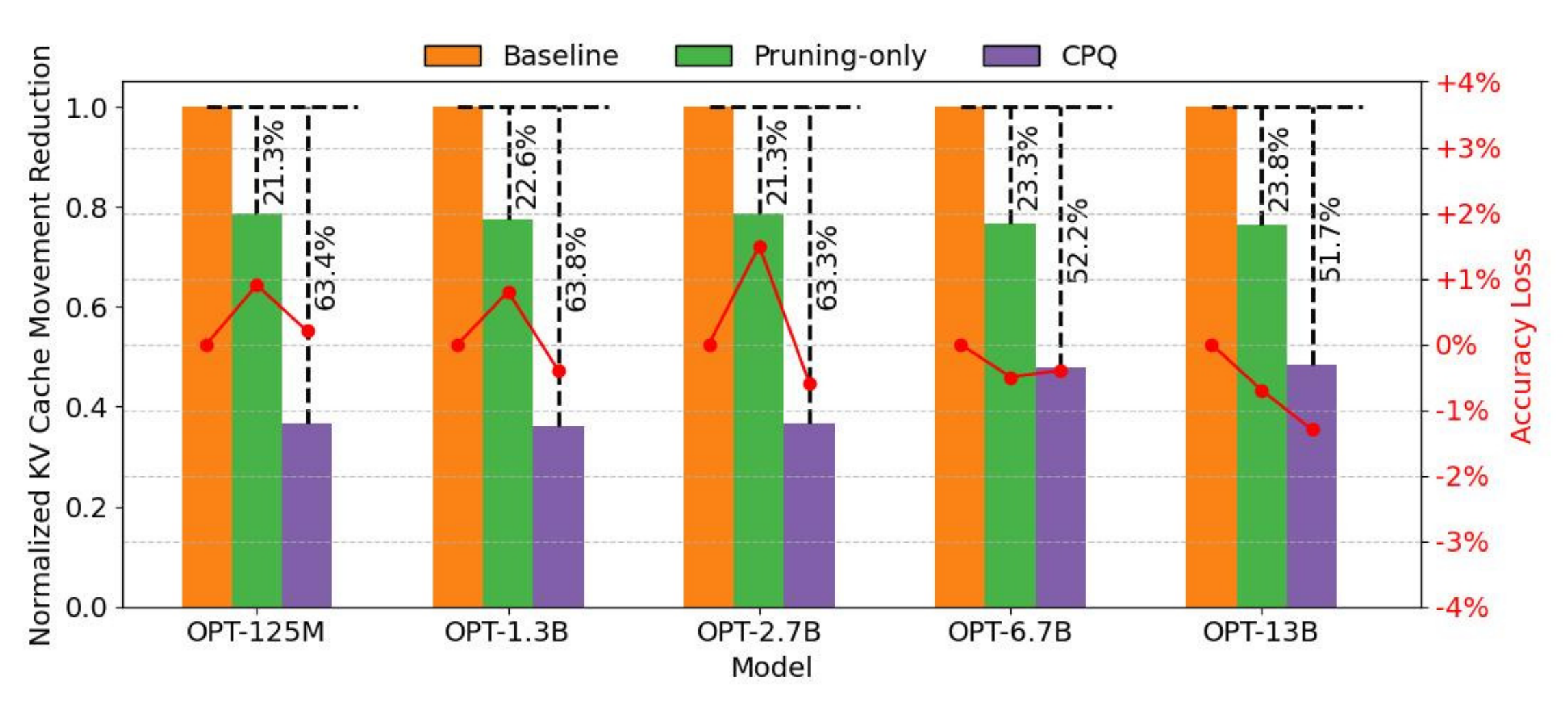}
    \vspace{-10pt}
    \caption{KV cache movement reduction using the CPQ.}
    \Description{fig11}
    \label{fig11}
    \vspace{-10pt}
\end{figure}

\textbf{Benefits of the CPQ Compression Method.} Figure~\ref{fig11} shows the KV cache movement reduction achieved by the CPQ compression method. The evaluation dataset is also LAMBADA. We use quantized W8A8 OPT models as the baseline. In the pruning-only approach, only the binary index and non-zero KV cache are transferred between the accelerator and off-chip memory. The optimal pruning configuration is obtained from the first stage of the DSE in Figure~\ref{fig10}. Unlike pruning-only, the CPQ transfers the index, label, and non-zero quantized KV cache, which can further reduce KV cache movement significantly. Compared to the baseline, the pruning-only and CPQ reduce KV cache movement by $22.3\%$ and $58.9\%$ on average, with almost no accuracy loss.

\subsection{Hardware Evaluation}

\begin{table}[tb]
  \centering
  \caption{Area and power breakdown for Titanus-125M core.}
  \label{table2}
  \vspace{-5pt}
  \begin{tabular}{@{}l c c c c@{}}
    \toprule
    \textbf{Modules} 
      & \textbf{Area[mm\textsuperscript{2}]} 
      & \textbf{\%} 
      & \textbf{Power[mW]} 
      & \textbf{\%} \\
    \midrule
    DCIM blocks\footnotemark[1]      
      & 80.5770  &  96.708  &  30,880.4604   &  96.652   \\
    CE       
      & 0.9904  &  1.189 &  100.8504  &  0.316  \\
    PU         
      & 0.0014  &  0.002 &  0.2147  &  0.001  \\
    QU         
      & 0.1901  &  0.228 &  51.7940  &  0.162  \\
    DQU                
      & 0.0042  &  0.005  &  1.5040  &  0.005  \\
    On-chip buffer   
      & 1.1952  &  1.435  &  893.6821  &  2.797  \\
    Others   
      & 0.3616  &  0.434  &  21.6439  &  0.067  \\
    \midrule
    \textbf{Total} 
      & \textbf{83.32} & \textbf{100} & \textbf{31,950.15} & \textbf{100} \\
    \bottomrule
  \end{tabular}
  \vspace{2pt}
  \parbox{\linewidth}{
    \footnotesize
    \footnotemark[1] The area and power of CIM macro are scaled from 22nm to 14nm~\cite{chih202116, wang2017energy}.
  }
  \vspace{-10pt}
\end{table}

\textbf{Area and Power Breakdown.} We provide five Titanus core designs to efficiently support the inference of different OPT models, namely Titanus-125M, Titanus-1.3B, Titanus-2.7B, Titanus-6.7B, and Titanus-13B. The main differences between these cores are the sizes of $Q_{cim}$, $K_{cim}$, $V_{cim}$, $Out_{cim}$, $FC1_{cim}$, and $FC2_{cim}$ blocks. We present the area and power breakdown of the Titanus-125M core in Table~\ref{table2}. The DCIM blocks are designed to store all static weights. Each CE has 4 VPUs, 16 MUs, and can perform 256 MACs simultaneously. The parallelism of PU, QU, and DQU is set to 16. The sizes of the Global buffer and SZ buffer are 4MB and 64KB, respectively. One Titanus-125M core has an area and power of $83.32 {mm}^2$ and $31.95W$, respectively. The DCIM blocks account for over $96\%$ of both area and power consumption. To support the proposed designs, the CE, PU, QU, and DQU modules contribute only $1.42\%$ of total area and $0.48\%$ of power.

\begin{figure}[tb]
    \centering
    \includegraphics[width=\linewidth]{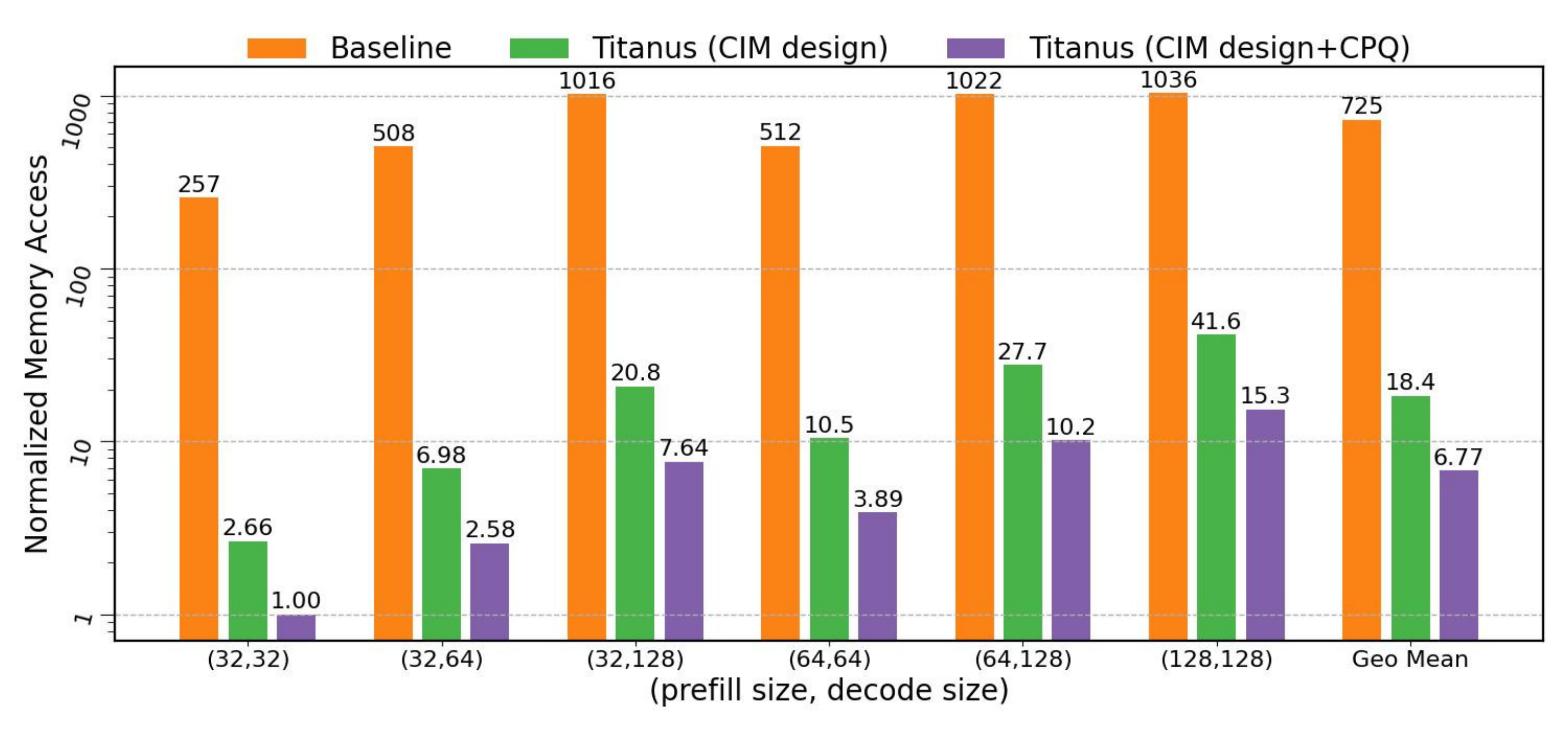}
    \vspace{-10pt}
    \caption{Memory access reduction of Titanus-125M design.}
    \Description{fig12}
    \label{fig12}
    \vspace{-10pt}
\end{figure}

\textbf{Data Movement Reduction.} To demonstrate the advantages of \textbf{Titanus}, we use 12 Titanus-125M cores~(OPT-125M with 12 decoder layers) to perform the end-to-end W8A8 OPT-125M inference under different combinations of prefill size and decode size~(\textit{batch size}$=$1). The baseline is GPUs and Transformer accelerators~\cite{zeng2024flightllm, tu202228nm1, wang2021spatten} that repeatedly reload all the LLM weights and KV cache during the decode stage. As shown in Figure~\ref{fig12}, compared to the baseline, \textbf{Titanus} with CIM design can reduce off-chip memory access by $39.4\times$ on average. \textbf{Titanus} stores all OPT-125M static weights in CIM blocks and only needs to reload the static weights once. With the CPQ compression method, \textbf{Titanus} achieves a further $107.1\times$ reduction in memory access due to the significantly less KV cache movement. Moreover, as the prefill size and decode size increase, the KV cache movement will dominate the overall off-chip memory access. Therefore, \textbf{Titanus} relies more on the CPQ method to reduce memory access in this case.

\begin{figure}[tb]
    \centering
    \includegraphics[width=\linewidth]{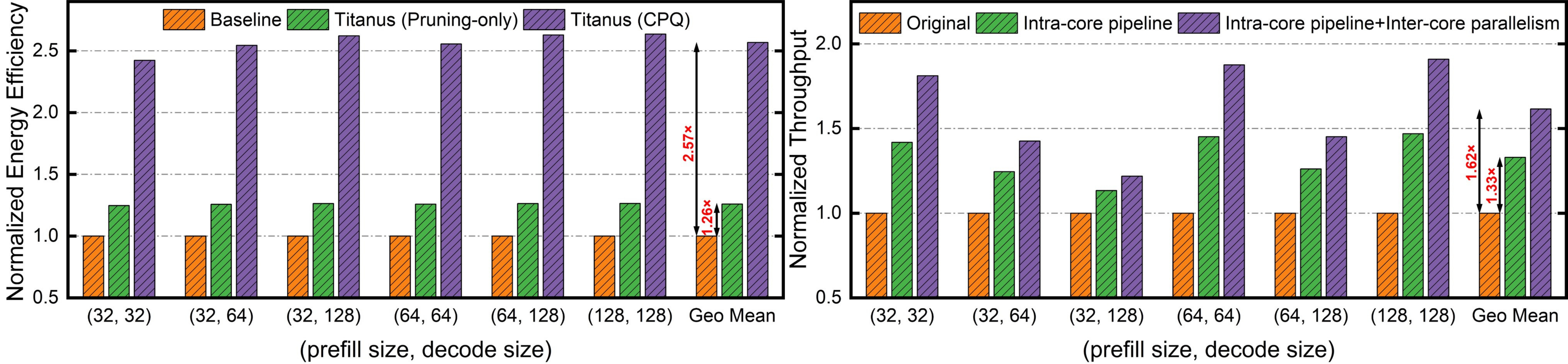}
    \vspace{-10pt}
    \caption{Energy efficiency~(Token/J) and throughput~(Token/s) improvement.}
    \Description{fig13}
    \label{fig13}
    \vspace{-10pt}
\end{figure}

\textbf{Energy Efficiency.} We set an ablation experiment to show the energy efficiency advantages of our CPQ compression method. The experiment configurations are consistent with those in Figure~\ref{fig12}. The baseline is the \textbf{Titanus} in which the PU, QU, and DQU modules are disabled. There are two additional configurations: one in which only the PU is enabled and another in which all three modules are enabled. As shown in the left part of Figure~\ref{fig13}, Titanus~(Pruning-only) achieves $1.26\times$ energy efficiency improvement on average compared to the baseline. This is because we only transfer the non-zero KV cache and utilize zeros from pruning to save energy. Furthermore, Titanus~(CPQ) only transfers the non-zero quantized KV cache and achieves an extra $2.04\times$ improvement.

\textbf{Throughput Improvement.} To show the effectiveness of our dataflow design, Figure~\ref{fig13}~(right) presents the throughput improvement brought by ICPI and ICPA. The experiment configurations are the same as those in Figure~\ref{fig12}. The original design in the figure represents \textbf{Titanus} does not adopt any method to optimize the time-to-first-token. We can observe that ICPI improves the throughput by $1.33\times$ on average. ICPA contributes to an extra $1.22\times$ improvement because the $i^{th}$ core does not need to wait for all prefill tokens to be fully processed in the $(i-1)^{th}$ core. Moreover, ICPI and ICPA exhibit higher throughput gains as the prefill size increases.

\begin{figure}[tb]
    \centering
    \includegraphics[width=\linewidth]{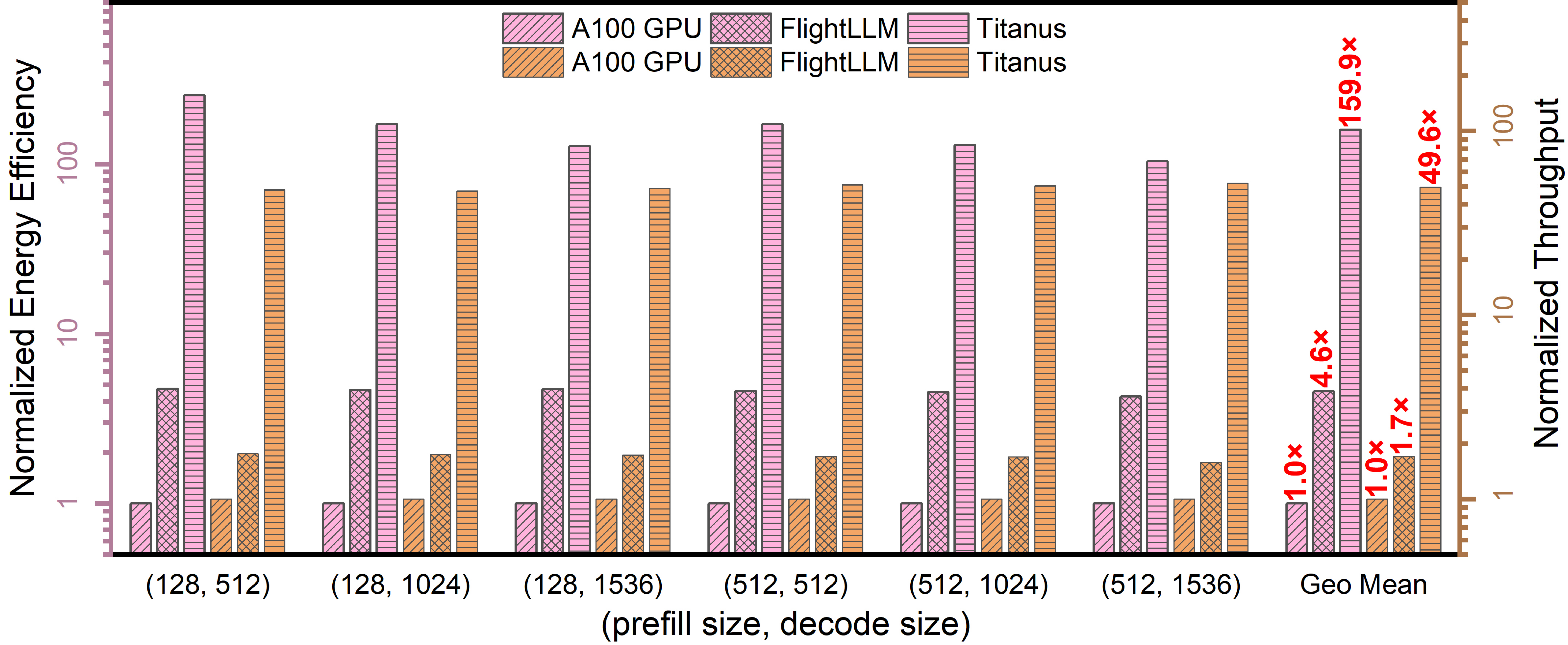}
    \vspace{-10pt}
    \caption{Energy efficiency and throughput of Titanus over A100 GPU and FlightLLM accelerator on OPT-6.7B.}
    \Description{fig14}
    \label{fig14}
    \vspace{-10pt}
\end{figure}

\textbf{Comparisons with GPU and SOTA Accelerator.} To make a fair comparison with FlightLLM~\cite{zeng2024flightllm}, we select the same model OPT-6.7B and evaluate its end-to-end performance on \textbf{Titanus}. We use the same six \textit{(prefill size, decode size)} combinations as~\cite{zeng2024flightllm} in our experiment. The OPT-6.7B running on A100 GPU is the original fp16 model. We use \textit{nvidia-smi} to measure the GPU power at runtime. The batch size is set as 1. As shown in Figure~\ref{fig14}, \textit{Titanus} achieves $159.9\times$~($49.6\times$) and $34.8\times$~($29.2\times$) energy efficiency~(throughput) compared to A100 GPU and FlightLLM respectively. The advantage of \textbf{Titanus} is mainly because it avoids the repeated movement of the LLM static weights while minimizing the movement of KV cache. Moreover, \textbf{Titanus} leverages CIM architecture to finish the large vector-matrix multiplication efficiently and utilizes the ICPI and ICPA dataflow to reduce the first token generation time.

\section{Conclusion}

We propose \textbf{Titanus}, a software-hardware co-design to compress the KV cache efficiently. We first propose the CPQ method to reduce KV cache movement. The HQE strategy is introduced to address the NiPCQ issue. To further reduce KV cache movement, we transfer only the non-zero quantized KV cache between the accelerator and off-chip memory. Moreover, we implement a two-stage DSE framework for our CPQ method. The ICPI and ICPA dataflow is designed to reduce the TTFT. Finally, \textbf{Titanus} achieves $159.9\times$~($49.6\times$) and $34.8\times$~($29.2\times$) energy efficiency~(throughput) compared to Nvidia A100 GPU and FlightLLM respectively.

\bibliographystyle{ACM-Reference-Format}
\bibliography{main}

\end{document}